\documentclass[%
 reprint,
 amsmath,amssymb,
aps,
prb,
]{revtex4-1}

\usepackage{graphicx}
\usepackage{dcolumn}
\usepackage{bm}
\usepackage{xcolor}

\newcommand*{\citen}{}
\DeclareRobustCommand*{\citen}[1]{%
	\begingroup
	\romannumeral-`\x 
	\setcitestyle{numbers}%
	\cite{#1}%
	\endgroup
}

\begin{document}
	
\preprint{APS/123-QED}

\title{Anisotropic magnons in a layered honeycomb ferromagnet}

\author{T.J.~Williams}
 \email{travis.williams@stfc.ac.uk}
 \affiliation{ISIS Pulsed Neutron \& Muon Source,
     STFC Rutherford Appleton Laboratory,
     Harwell Campus, Didcot OX11 0QX, United Kingdom}
\author{D.L. Abernathy}
\author{M.D.~Lumsden}
\affiliation{Neutron Scattering Division,
    Neutron Sciences Directorate,
    Oak Ridge National Lab,
    Oak Ridge, TN, 37831, USA}
\author{J.-Q.~Yan}
\author{A.D. Christianson}
\affiliation{Materials Science \& Technology Division,
    Physical Sciences Directorate,
    Oak Ridge National Lab,
    Oak Ridge, TN, 37831, USA}

\date{\today}

\begin{abstract}
Recent experimental and theoretical studies have suggested a possible Dirac magnon gap in the two-dimensional ferromagnetic semiconductor CrSiTe$_3$.  Detailed neutron scattering measurements were performed to shed light on the existence of the magnon gap, and suggest that the gap is very small or non-existent, with previous measurements being complicated by experimental factors.  During these measurements, it was found that the out-of-plane couplings could explain the usual property of the increase in the magnetic transition temperature when CrSiTe$_3$ is exfoliated to monolayers.  Furthermore, the material was shown to have anisotropic magnons along the out-of-plane direction, through the proposed Dirac point.  We speculate that this is due to an exchange anisotropy, though Kitaev-like interactions alone cannot explain the spectra. 

\begin{description}
\item[PACS numbers]{75.30.Ds, 75.40.-s, 75.50.Pp, 85.75.-d}
\end{description}
\end{abstract}

\maketitle

\section{\label{sec:level1}Introduction}

Two-dimensional (2D) magnetic materials have recently emerged as a vibrant platform for exploring novel quantum phenomena and for advancing next-generation spintronic devices~\cite{Bhimanapati_15,Milosevic_21}. Among these materials, chromium-based trichalcogenides, such as CrSiTe$_3$, have attracted particular interest due to their layered van der Waals structure, intrinsic ferromagnetism, and strong spin–orbit coupling~\cite{Liu_23,Casto_15}. CrSiTe$_3$ is a semiconducting ferromagnet that retains long-range magnetic order down to the monolayer limit, making it a promising candidate for studying low-dimensional magnetism and collective spin excitations. Among the promising properties of CrSiTe$_3$ is the possibility of Dirac magnons~\cite{Pershoguba_18}.  Analogous to Dirac fermions in graphene, Dirac magnons—represent a topologically nontrivial excitation characterized by linear band crossings at discrete momentum points in the Brillouin zone. The existence of Dirac magnons in honeycomb-lattice magnets opens up new avenues for exploring bosonic analogs of topological phases, including magnonic analogues of the quantum Hall and spin Hall effects~\cite{Lu_21}. The identification and control of such excitations in real materials is a critical step toward realizing topological magnonics, where robust, low-dissipation spin transport could be achieved via edge modes protected by symmetry.

In this work, we re-investigate the magnons of CrSiTe$_3$, focusing on the possible emergence of Dirac magnons due to its honeycomb-like arrangement of Cr atoms and underlying magnetic interactions. This sheds light on the role of anisotropy, exchange coupling, and Dzyaloshinskii–Moriya (DM) interactions in shaping the topological character of the magnon bands.

Previous work by our group on single crystals of CrSiTe$_3$ have shown a small distortion of the Te octahedra along the $c$-axis that result in a small easy-axis anisotropy~\cite{Casto_15}.  This was confirmed using elastic and inelastic neutron scattering measurements on single crystals of CrSiTe$_3$ that found a small spin gap of 0.075~meV that gives rise to the ferromagnetic ground state, with moments oriented parallel to $c$~\cite{Williams_15}.  The temperature-dependence of the magnetic Bragg peak shows a critical exponent that is very close to the 2D Ising value, and diffuse scattering around the magnetic Bragg peaks provide further evidence of quasi-2D behavior that persists well above the ordering temperature.  Time-of-flight (TOF) and triple-axis inelastic neutron scattering measured performed on single crystals aligned in the $[H~0~L]$ scattering plane~\cite{Williams_15} demonstrate well-ordered spin waves below T$_C$, with in-plane diffuse spin correlations persisting up to room temperature~\cite{Williams_15}.  Modeling and fits of the spin waves were done with a Hamiltonian that included 3 Heisenberg in-plane exchange terms and 1 exchange along the $c$-axis, all of which were found to be ferromagnetic, in addition to the single ion term~\cite{Williams_15}.  This work showed that the moments were only weakly coupled along the $c$-axis, with the dominant exchange terms lying in the plane, consistent with the diffuse elastic and inelastic scattering that showed correlations within the $ab$-plane that persist well above T$_C$.

More recent neutron spectroscopy measurements have suggested the existence of a Dirac magnon gap at the $K$-point in CrSiTe$_3$~\cite{Zhu_21}, which would give rise to corresponding in-gap topological edge states.  This gap was attributed to the existence of a Dzyaloshinskii-Moriya (DM) interaction of 0.12meV between magnetic ions with broken inversion symmetry, while the remaining terms in the Hamiltonian comprised two in-plane exchange terms, $J_{ab1}$ and $J_{ab2}$, and two out-of-plane terms, $J_{c1}$ and $J_{c2}$. The differences in the Hamiltonian and the sizable DM term suggested a re-examination of our previous neutron data, as well as the need to perform additional measurements in the $[H~H~L]$ scattering plane to examine the magnon dispersions around the Dirac ($K$) point.

The work in Ref.~\citen{Zhu_21} did not describe any corrections due to instrumental resolution, which may alter the size of the gap and thus the value of the DM term.  Additionally, the triple-axis measurements in that work did not account for a $k_i \rightarrow k_i$ spurion arising from the (0~0~3) nuclear Bragg peak.  This is an unusual spurion, as it involves coherent scattering from the monochromator and sample, but incoherent scattering from the PG analyzer. Although incoherent scattering from PG is very weak, the (0~0~3) Bragg peak is the most intense nuclear peak, and this spurion should be of the same order of magnitude as the intensity of the inelastic scattering from the magnons.  On a triple axis instrument, this spurion appears at (0.3~0.3~6) with an energy transfer of 8~meV, which is very close to the location of the proposed Dirac magnon gap.  This provided additional motivation to perform measurements that could investigate this as a source of error in measuring the magnitude of the gap.

\section{\label{sec:level2}Experimental Details}

CrSiTe$_3$ single crystals were grown using a self-flux technique, as previously reported~\cite{Casto_15}.  CrSiTe$_3$ is rhombohedral, crystallizing in the space group $R \overline{3}$.  Susceptibility and neutron diffraction measurements showed a ferromagnetic transition at T$_{C}$~=~33(1)~K~\cite{Casto_15,Williams_15}, in agreement with other reports~\cite{Carteaux_95,Yang_23}.  In order to study the existence of a gap in the magnon spectrum and the impact of experimental resolution, inelastic neutron scattering measurements were performed at the HB-3 thermal triple axis spectrometer of the High-Flux Isotope Reactor, as well as the ARCS time-of-flight spectrometer at the Spallation Neutron Source of the Oak Ridge National Laboratory.

A single crystal of mass 3.7~g was aligned with a [$H$~$H$~$L$] scattering plane for the ARCS and HB3 measurements.  The ARCS measurements were performed in a closed-cycle refrigerator with a base temperature of 4.0~K.  Measurements were performed in a high-flux configuration with fixed incident energies of 30~meV and 15~meV.  The HB3 measurements were also performed in a closed-cycle refrigerator with a base temperature of 4.0~K using fixed final energies of 14.7~meV and 5.5~meV. PG(002) monochromator and analyzer crystals were used with PG filters, and the collimation was 48'-40'-40'-120'.  Where mentioned, our previous measurements on CrSiTe$_3$ aligned in the [$H$~0~$L$] scattering plane were reported in Ref.~\citen{Williams_15}.

\section{\label{sec:level3}Dirac Magnon Gap}

To investigate the nature of a Dirac magnon gap at the $K$-point, measurements were performed on the ARCS thermal time-of-flight spectrometer.  Following the work of Ref.~\citen{Zhu_21}, which ascribed the magnon gap to a second nearest-neighbor DM interaction, our measurements were performed with the crystal aligned in the [$H$~$H$~$L$] scattering plane, focused on reciprocal space directions of (\(\frac{1}{3}\)~\(\frac{1}{3}\)~$L$), which passes through the $K$-point, and  (\(\frac{1}{2}\)~\(\frac{1}{2}\)~$L$), where the value of the DM interaction does not contribute to the magnon dispersion.  This allows for a separation of the Heisenberg interactions along the $c$-axis with a possible DM interaction.   

The measured bandwidth along (\(\frac{1}{2}\)~\(\frac{1}{2}\)~$L$) was used to determine the exchange interactions along the $c$-axis. this was measured to be 0.61~meV, slightly smaller than 0.73~meV from our previous measurement~\cite{Williams_15}, taken along a different reciprocal space direction. Using the single ion anisotropy, $D_{ij}~=~0.0252$~meV and the geometry of the $R$\={3} stacking, the relationship between $J_{c1}$ and $J_{c2}$ can be constrained.  There is only 1 nearest neighbour ($J_{c1}$) bond along $c$, but there are 6 $J_{c2}$ bonds.  When combined with the phase factor of 1.5 and fitting to the measured bandwidth, we can constrain $J_{c1}+9 J_{c2} = -0.61$~meV. Additional interactions out of the plane such as $J_{c3}$ result in over-parametrising the data and do not provide a meaningful improvement in the fit. A fit to the dispersion gives the value of these interactions listed in Table~\ref{exch_caxis}.

The opposite sign of the two interactions indicates significant frustration in bulk samples along the $c$-axis, while the larger number of J$_{c2}$ bonds relative to J$_{c1}$ means that the ferromagnetic interaction dominates and gives rise to the ferromagnetic ground state.  This difference in the sign of the interaction also offers an explanation for the low $T_C$ considering the intraplane interactions that persist up to room temperature~\cite{Williams_15} and that the ordering transition of CrSiTe$_3$ increases as the material is exfoliated to thin films and monolayers~\cite{Lin_16}, relieving the long-range interlayer frustration.  While the nearest-neighbor interaction along the $c$-axis is antiferromagnetic, the $J_{c2}$ interaction is of comparable strength and the larger multiplicity stabilizes the 3D ferromagnetism in the material.  This has implications for the nature of the ordering of the material in the presence of lattice distortions, in agreement with theoretical predictions for CrSiTe$_3$ under in-plane strain~\cite{Sivadas_15}.  Devices and heterostructures where CrSiTe$_3$ is placed on a substrate will similarly be subjected to lattice strain that can affect the degree of magnetic frustration~\cite{Pei_18}.

\begin{figure}[tbh]
\begin{center}
\includegraphics[angle=0,width=\columnwidth]{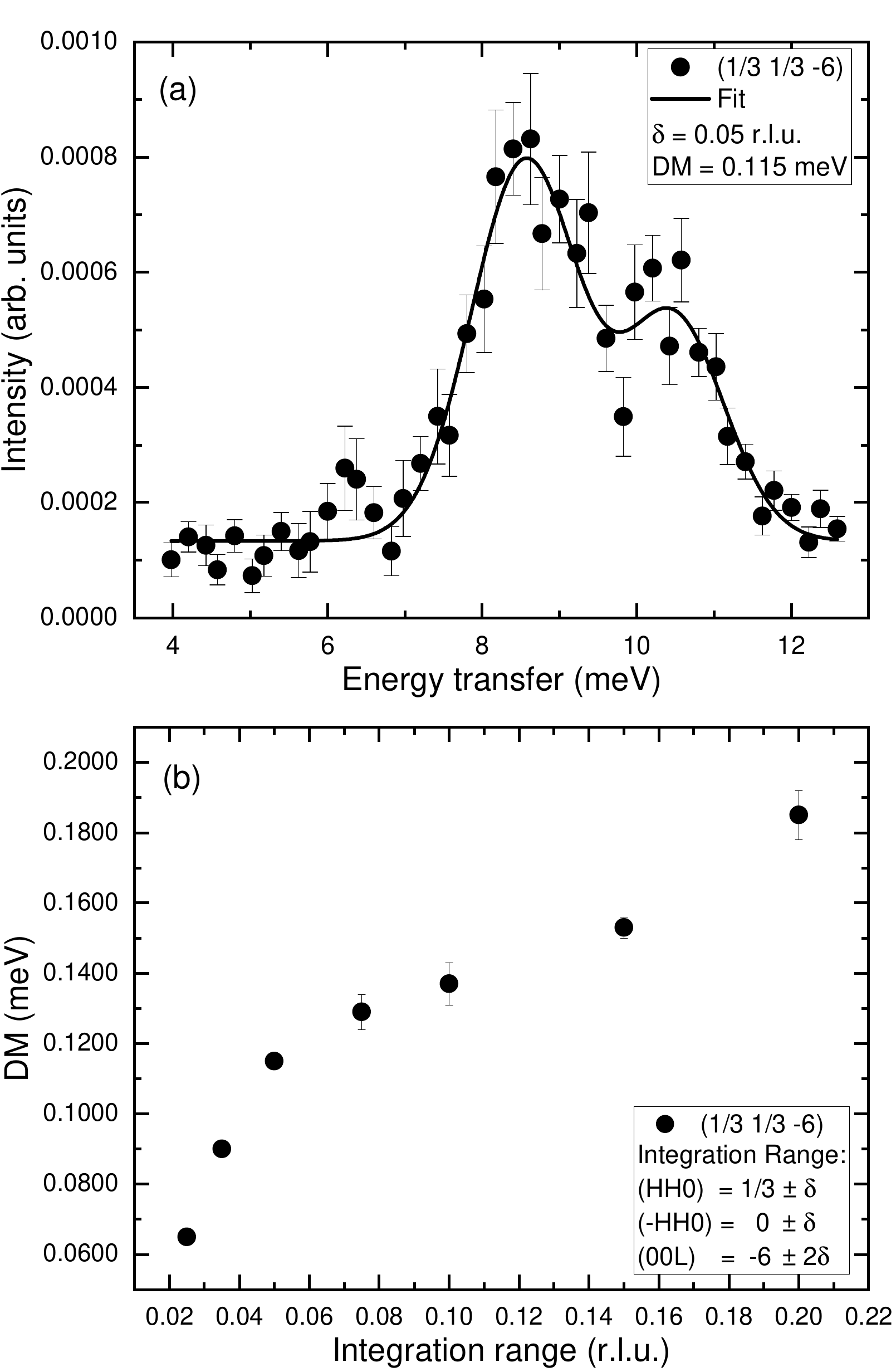}
\caption{\label{ARCS} (a) A line cut through the $K$-point by integrating (\(\frac{1}{3}\)$\pm$0.05~\(\frac{1}{3}\)$\pm$0.05~-6$\pm$0.1) and $\Delta E$~=~0.25~meV using the data collected on ARCS.  The data was then fit to the Hamiltonian descibed in the text, including the DM term.  The fit value for the DM term was 0.115(3)~meV, in good agreement with Ref.~\citen{Zhu_21}. (b) When the procedure done in the top panel was repeated for integrating different $Q$-volumes around the $K$-point, the fit value of the DM term was found to monotonically decrease with the integration range.  The fits did not converge for volumes with $\delta <$~0.025 r.l.u.}
\end{center}
\end{figure}

To look for the presence of a magnon gap, we looked at the constant-$\vec{Q}$ cut at the $K$-point = (\(\frac{1}{3}\)~\(\frac{1}{3}\)~-6), shown in Fig.~\ref{ARCS}(a).  The spurion described above will not be present in time-of-flight data, but it is subject to other resolution effects, as described below.  To create this cut, the data was integrated to cover the volume of $\vec{Q}$-space contained in (\(\frac{1}{3}\)+$\delta$,~\(\frac{1}{3}\)+$\delta$,~-6+2$\delta$) with $\delta$=0.05~r.l.u.  The data was fit to the following Hamiltonian:

\begin{equation}
\begin{split}
H&= J_{ab1} \sum_{i}{\vec{S}_i \cdot \vec{S}_{i+1}} + 
J_{ab2} \sum_{i}{\vec{S}_i \cdot \vec{S}_{i+2}} \\
&+ J_{c1} \sum_{i}{\vec{S}_i \cdot \vec{S}_{i+3}} + 
J_{ab3} \sum_{i}{\vec{S}_i \cdot \vec{S}_{i+4}} \\
&+ J_{c2} \sum_{i}{\vec{S}_i \cdot \vec{S}_{i+5}} - 
D_{ij} \sum_{i<j}{\vec{S}_i \times \vec{S}_j} 
\label{eq1}
\end{split}
\end{equation}

where the exchanges out to the 6$^{th}$ nearest neighbor are included.  This Hamiltonian follows work of Ref.~\citen{Williams_15}, where a 3$^{rd}$ nearest neighbor term, $J_{ab3}$, was needed to capture the spin wave dispersions in the plane, and the work of Ref.~\citen{Zhu_21}, where a 2$^{nd}$ out-of-plane exchange, $J_{c2}$, and the DM term, $D_{ij}$, was included.

In order to fit the cut in Fig.~\ref{ARCS}(a) to Eq.~\ref{eq1}, the values for J$_{c1}$ and J$_{c2}$ obtained above were used, along with the values obtained from our previous work\cite{Williams_15}. The value of the nearest-neighbour interaction was fit along with the DM term, and this gave a value for the DM interaction of 0.115(3)~meV.  This is in excellent agreement with previous work in Ref.~\citen{Zhu_21}, which found a value of $D$~=~0.12~meV.  The values of the parameters are summarised in Table~\ref{exch_caxis}.  As we discuss in the following paragraph, this demonstrates reproducibility of the data, but the existence of a DM interaction may be more complex.

\begin{table}[htb]
\begin{center}
\begin{tabular}{|l|l|l|l|l|}
\hline
~Exchange~ & ~\# Bonds~ & ~~Vector~~ & ~Distance~ & ~Value (meV)~
  \\
\hline
~~~~~$J_{ab1}$ & ~~~~~~3 & ~~(\(\frac{1}{3}\)~~\(\frac{2}{3}\)~~0)~ & ~~3.907 \AA & ~~~~-1.63(33) \\
~~~~~$J_{ab2}$ & ~~~~~~6 & ~~(1~~0~~0)~ & ~~6.768 \AA & ~~~~-0.010(-) \\
~~~~~$J_{ab3}$ & ~~~~~~3 & ~~(\(\frac{2}{3}\)~\(-\frac{2}{3}\)~~0)~ & ~~7.814 \AA & ~~~~-0.285(-) \\
~~~~~$J_{c1}$ & ~~~~~~1 & ~~(0~~0~~\(\frac{1}{3}\))~ & ~~7.007 \AA & ~~~~0.080(9) \\
~~~~~$J_{c2}$ & ~~~~~~6 & ~(\(\frac{2}{3}\)~~\(\frac{1}{3}\)~~\(\frac{1}{3}\)) & ~~8.700 \AA & ~~~-0.077(2) \\
\hline
\end{tabular}
\end{center}
\caption[]{The out-of-plane exchange constants calculated from the neutron scattering data measured along (\(\frac{1}{2}\)~\(\frac{1}{2}\)~$L$).  While the nearest-neighbor exchange is antiferromagnetic, the next-nearest neighbor exchanges dominates due to a greater number of J$_{c2}$ bonds.  This results in the observed ferromagnetic ground state, while the competition from these two interactions may drive the reduction in T$_C$ from the monolayer.}
\label{exch_caxis}
\end{table}

Recent work on another Cr honeycomb material, CrI$_3$, has demonstrated that an apparent magnon gap was likely overestimated due to finite instrumental resolution and the data integration volume used in the analysis~\cite{Do_22}.  To perform a similar analysis on CrSiTe$_3$, the integration volume defined by $\delta$ was decreased from 0.2~r.l.u. and fitting the data in the same manner as described above to obtain a value for the DM interaction.  The results are shown in Fig.~\ref{ARCS}(b), and show that the resulting value of the DM interaction decreases monotonically with the integration volume, as was observed in CrI$_3$.  For values of $\delta<$~0.025~r.l.u., the fits did not converge, so it is unclear whether the DM term vanishes or is present, but much smaller than reported.  

\section{\label{sec:level4}Instrumental Resolution}

\begin{figure}[tbh]
\begin{center}
\includegraphics[angle=0,width=\columnwidth]{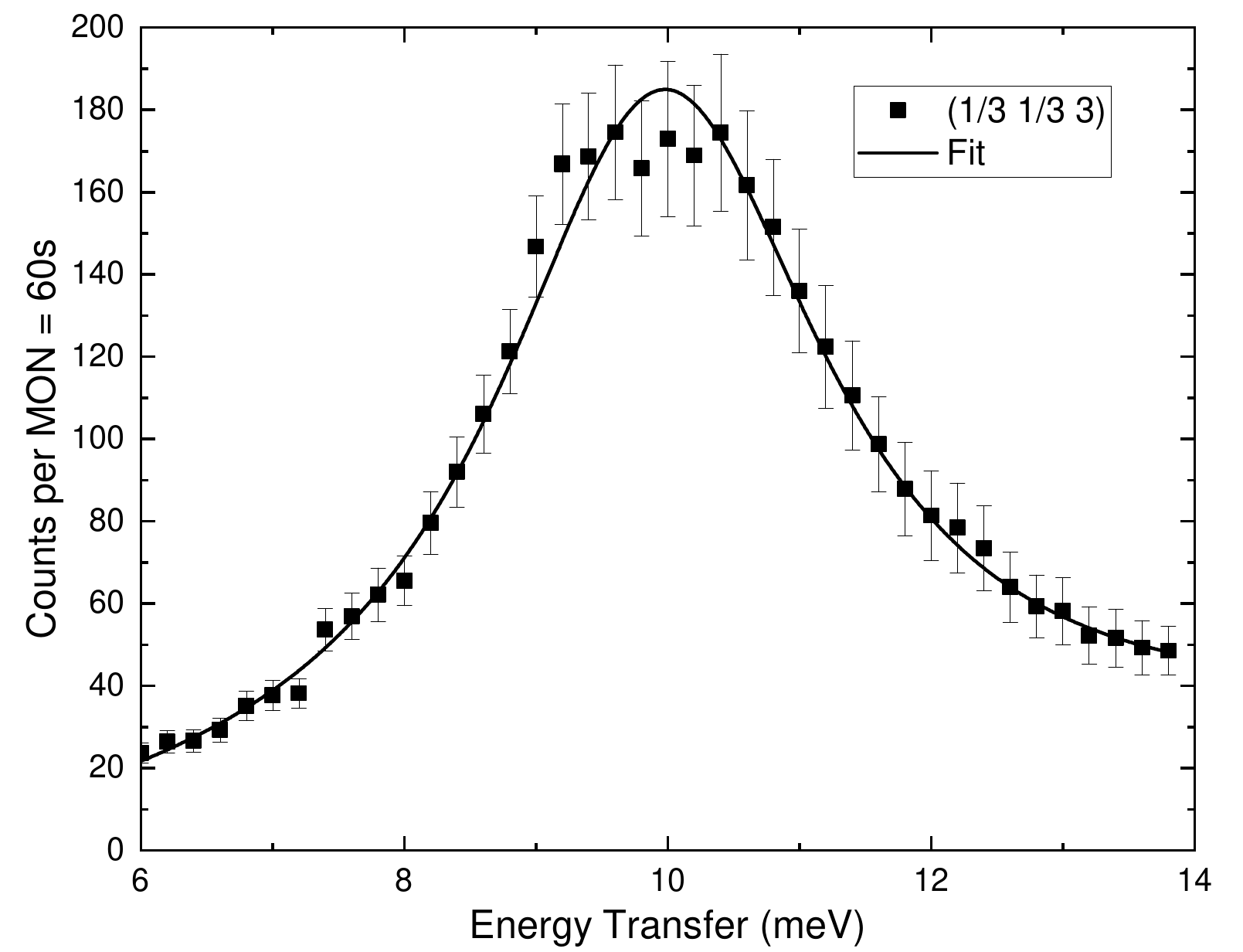}
\caption{\label{hb3} A constant-$Q$ scan at the $K$-point on the HB3 thermal triple axis spectrometer.  A background scan was also performed using the same configuration, but driving the analyzer flat.  This removes coherent scattering from the analyzer, but preserves the incoherent signal that is a source of spurious scattering.  The background-subtracted data is shown, where the solid line is a fit to the Hamiltonian with zero DM term, and convoluted with the instrumental resolution.}
\end{center}
\end{figure}

To make more quantitative conclusions about the DM interaction and the effect of instrumental resolution, measurements were also performed on the thermal triple axis spectrometer HB3.  Constant-$\vec{Q}$ measurements at the $K$-point (\(\frac{1}{3}\)~\(\frac{1}{3}\)~3) were performed with fixed final energies of $E_f$~=~5.5~meV and 14.7~meV.  To remove spurious scattering and other sources of background, a background was measured by driving the PG analyzer to zero scattering angle and performing the identical constant-$\vec{Q}$ scan.  This removes coherent scattering from the analyzer but preserves the inchorent scattering that gives rise to the spurion at (0.3,~0.3,~3) and 8~meV.  The background-subtracted data was fit to the Hamiltonian described in Eq.~\ref{eq1}, with the $J_{c1}$ and $J_{c2}$ values determined earlier and the DM term fixed to zero.  This Hamiltonian was convoluted with the instrumental resolution calculated with RESLIB~\cite{RESLIB}, and the corresponding fit is shown in Fig.~\ref{hb3}.

The data is seen to be consistent with the absence of a DM interaction, though it does not rule out a small term.  Fitting this data with a non-zero DM term did not converge due to over-parameterization from the in-plane interactions, but it was seen that the data was not consistent with a DM term of the magnitude reported in Ref.~\citen{Zhu_21}.  This suggests that the behavior of CrSiTe$_3$ is likely similar to that of CrI$_3$~\cite{Do_22}, CrBr$_3$~\cite{Nikitin_22}, CrCl$_3$~\cite{Chen_21} and other related materials that may possess a very small gap.  The nature of the interactions is more complicated than the Hamiltonian used, as it does not account for the anisotropic magnons that were observed, descirbed in the Section below.

\section{\label{sec:level5}Anisotropic Magnons}

\begin{figure*}[tbh]
\begin{center}
\includegraphics[angle=0,width=7in]{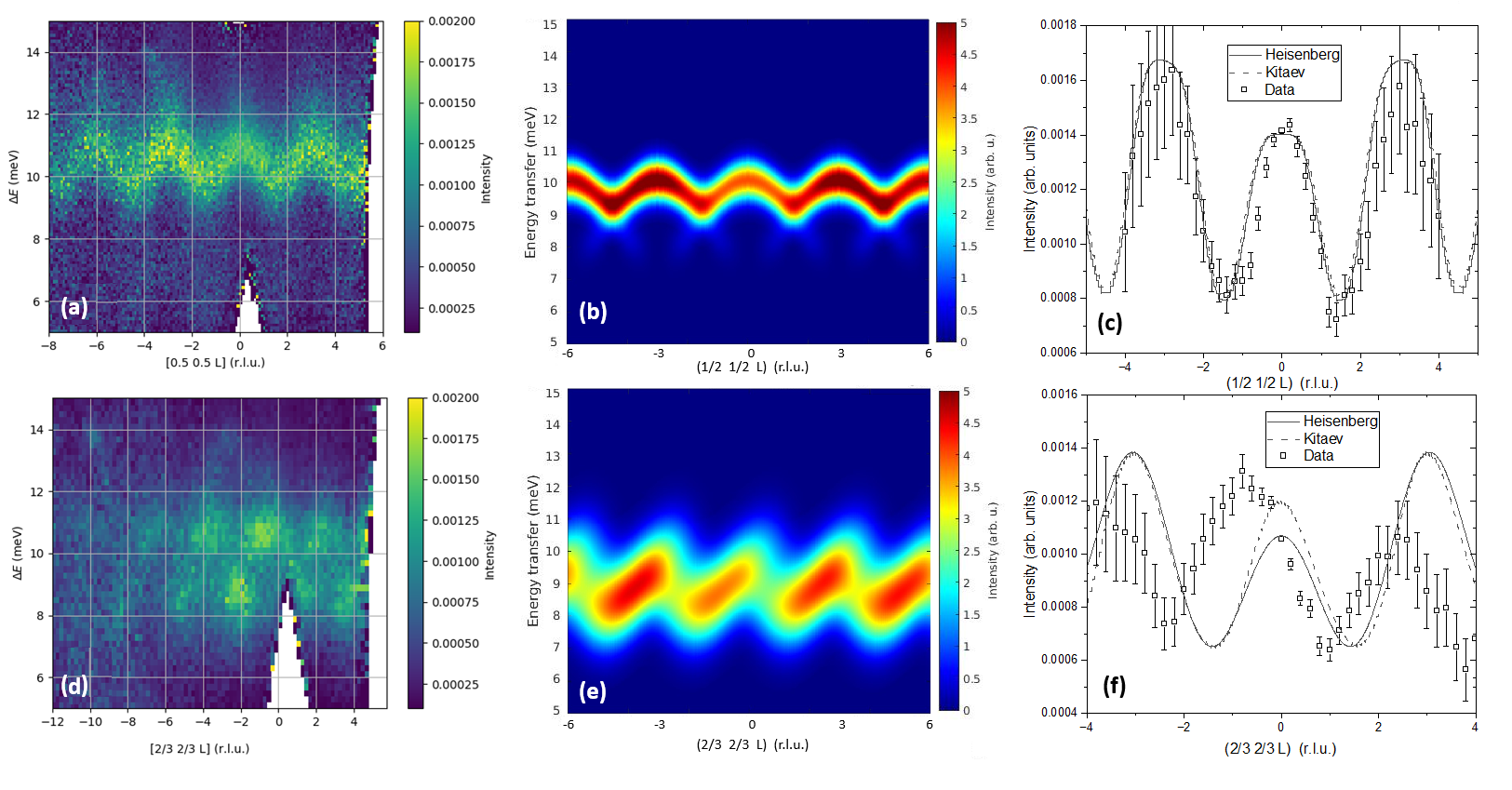}
\caption{\label{spinw} (color online) (a) Spin waves measured along the (\(\frac{1}{2}\)~\(\frac{1}{2}\)~$L$) direction, where there is no contribution from a DM term.  This data was used to constrain the fits to the out-of-plane excahnge constants.  (b) Spin wave simulations of the (\(\frac{1}{2}\)~\(\frac{1}{2}\)~$L$) dispersion with the Kitaev interaction described in the text.  (c) The data along (\(\frac{1}{2}\)~\(\frac{1}{2}\)~$L$) at the top of the band ($\Delta E$~=~10.5~meV) overlaid with simulations using a purely Heisenberg Hamiltonian and no DM interaction (solid line), and the same Hamiltionian, but with the nearest-neighbor in-plane interaction being a Kitaev interaction (dashed line).  Both agree well with the data, as this reciprocal space direction is not sensitive to changes in these parameters.  (d) Spin waves measured along the (\(\frac{2}{3}\)~\(\frac{2}{3}\)~$L$) direction, where the dispersion is seen to be anisotropic relative to $L$~=~0. (e) The simulations along the Fig.~\ref{spinw}(a) direction including the Kitaev interaction.  This distorts the spin wave spectrum and produces an anisotropy along the $L$ direction. (f) The data overlaid with the Heisenberg and Kitaev exchange simulations along the top of the band ($\Delta E$~=~10~meV).  This shows that the anistropy produced by the Kitaev term does not change the band minima/maxima, fundamentally different from the spin wave anisotropy observed in the data.}
\end{center}
\end{figure*}

The ARCS data showing the dispersions along the (\(\frac{1}{2}\)~\(\frac{1}{2}\)~$L$) direction is shown in Fig.~\ref{spinw}(a).  Simulation of this spin wave spectrum were done using SpinW~\cite{SPIN_W} utilising the Hamiltonian described above.  Along this direction, the spin wave spectrum is not sensitive to the presence of a DM interaction, though for these calculations the DM interaction is fixed to zero. The data for the (\(\frac{2}{3}\)~\(\frac{2}{3}\)~$L$) direction is shown in panel (d).  It was observed that the spectrum along this direction is not symmetric about $L$=0 ($L$ is not equivalent to -$L$), while the simulated data was symmetric.

This anisotropic magnon dispersion is allowed by the crystal symmetry, since there is no mirror plane along the $c$-axis.  However, the anisotropy cannot be reproduced with the linear spin wave theory (LSWT) calculations in SpinW, including the introduction of a DM interaction.  Using the software package Sunny~\cite{SUNNY}, the introduction of a Kitaev interaction along the nearest-neighbor in-plane exchange ($J_{ab1}$) does not change the symmetry of the bands, but it does produce some anisotropy in the spin wave intensity.  This anisotropy appears in the middle of the band, which does not match the experimental data.  In the data, the intensity remains at the extrema of the band, where the data are anisotropic, but the model is not.  Fig.~\ref{spinw}(b) and (e) show the simulated data along these directions with the inclusion of this Kitaev interaction.  To highlight the difference between the data and simulations, Fig.~\ref{spinw}(c) and (f) show line cuts along these two directions at the top of the spin wave band.  This shows that for the (\(\frac{2}{3}\)~\(\frac{2}{3}\)~$L$) direction, the introduction of the anisotropic Kitaev exchange does not change the $L$-symmetry of the band maxima and minima.  Other exchanges, including in-plane DM interactions, biquadratic exchange, and other Kitaev terms, were introduced to attempt to explain the data, but could not explain the observed asymmetric magnons.  It may indicate some other form of exchange anisotropy is responsible for this behavior, but this is an open question.

This anisotropy is not present along any other reciprocal space direction measured.  It is also an open question as to why the $K$-$H$-$K$' direction is unique in this regard, especially when compared to other reciprocal space directions, such as $\Gamma$-$A$-$\Gamma$'.

\section{\label{sec:level6}Conclusions}

We have re-examined the magnon dispersions in CrSiTe$_3$ in light of the possible emergence of Dirac magnons. Neutron spectroscopy measurements have demonstrated the importance of the 2$^{nd}$ nearest neighbor out-of-plane interaction, $J_{c2}$, in stabilizing the long-range order.  The frustration of $J_{c1}$ and $J_{c2}$ may explain the increase in the magnetic transition temperature when CrSiTe$_3$ is exfoliated to monolayers or is subjected to in-plane strain.  A careful study of the magnon dispersion around the $K$-point suggests that the observed gap appears to be enhanced due to resolution effects and spurious scattering.  If there is a DM term present in this material, it is likely much smaller than previously reported.  

Finally, an anisotropic magnon dispersion was observed along the (\(\frac{2}{3}\)~\(\frac{2}{3}\)~$L$) direction, which encompasses the $K$-point, with no anisotropy observed for other reciprocal lattice directions.  This anisotropy has not been explained with Kitaev interactions, but simulations of the data using a Kitaev term have shown anisotropic spectral weight, suggesting that the anisotropy observed in the data may arise due to other anisotropic exchange.  Determining the form of this exchange interaction may shed further light on the Hamiltonian and the nature of any gapped excitations or Dirac magnons.

\section{\label{sec:level7}Acknowledgments}

This research at ORNL's High Flux Isotope Reactor and Spallation Neutron Source was 
sponsored by the Scientific User Facilities Division, Office of Basic Energy 
Sciences, U.S. Department of Energy. The work of ADC was was supported by the U.S. Department
of Energy, Office of Science, Basic Energy Sciences, Materials Science and Engineering Division.

This manuscript has been authored by UT-Battelle, LLC under Contract No. DE-AC05-00OR22725 with the U.S. Department of Energy.  The United States Government retains and the publisher, by accepting the article for publication, acknowledges that the United States Government retains a non-exclusive, paid-up, irrevocable, world-wide license to publish or reproduce the published form of this manuscript, or allow others to do so, for United States Government purposes.  The Department of Energy will provide public access to these results of federally sponsored research in accordance with the DOE Public Access Plan (http://energy.gov/downloads/doe-public-access-plan).

\section{\label{sec:level8}Data Availability}

All data supporting this manuscript are publicly available~\cite{Data}.


\begin{thebibliography}{1}

\bibitem{Bhimanapati_15}G.R.~Bhimanapati, Z.~Lin, V.~Meunier, Y.-W.~Jung, J.J.~Cha, S.~Das, D.~Xiao, Y.-W.~Son, M.S.~Strano, V.R.~Cooper, L.~Liang, S.G.~Louie, E.~Ringe, W.~Zhou, B.G.~Sumpter, H.~Terrones, F.~Xia, Y.~Wang, J.~Zhu, D.~Akinwande, N.~Alem, J.A.~Schuller, R.E.~Schaak, M.~Terrones and J.A.~Robinson, Recent Advances in Two-Dimensional Materials beyond Graphene. {\em ACS Nano}. {\bf 9}, 11509 (2015).

\bibitem{Milosevic_21}M.V.~Milo\u{s}evi\'{c} and D.~Mandrus, 2D Quantum materials: Magnetism and superconductivity. {\em J. Appl. Phys.} {\bf 130}, 180401 (2021).

\bibitem{Liu_23}P.~Liu, Y.~Zhang, K.~Li, Y.~Li and Y.~Pu, Recent advances in 2D van der Waals magnets: Detection, modulation, and applications. {\em IScience}. {\bf 29}, 107584 
(2023).

\bibitem{Casto_15}L.~Casto, A.J.~Clune, M.O.~Yokosuk, J.L.~Musfeldt, 
T.J.~Williams, H.L.Zhuang, M.W.~Lin, K.~Xiao, R.G.~Hennig, B.C.~Sales, 
J.-Q.~Yan and D.~Mandrus,S trong spin-lattice coupling in CrSiTe$_3$. {\em APL Mat.} {\bf 3}, 041515 (2015).

\bibitem{Pershoguba_18}S.S.~Pershoguba, S.~Banerjee, J.C.~Lashley, J.~Park, H.~\AA gren, G.~Aeppli and A.V.~Balatsky, Dirac Magnons in Honeycomb Ferromagnets. {\em Phys. Rev. X.} {\bf 8}, 011010 (2018).

\bibitem{Lu_21}Y.-S.~Lu, J.-L.~Li and C.-T.~Wu, Topological Phase Transitions of Dirac Magnons in Honeycomb Ferromagnets. {\em Phys. Rev. Lett}. {\bf 127}, 217202 (2021).

\bibitem{Williams_15}T.J.~Williams, A.A.~Aczel, M.D.~Lumsden, S.E.~Nagler, M.B.~Stone, J.-Q.~Yan and D.~Mandrus, Magnetic correlations in the quasi-two-dimensional semiconducting ferromagnet CrSiTe$_3$. {\em Phys. Rev. B.} {\bf 92}, 144404 (2015).

\bibitem{Zhu_21}F.~Zhu, L.~Zhang, X.~Wang, F.J.~dos Santos, J.~Song, T.~Mueller, K.~Schmalzl, W.F.~Schmidt, A.~Ivanov, J.T.~Park, J.~Xu, J.~Ma, S.~Lounis, S.~Blügel, Y.~Mokrousov, Y.~Su and T.~Brückel, Topological magnon insulators in two-dimensional van der Waals ferromagnets CrSiTe$_3$ and CrGeTe$_3$: Toward intrinsic gap-tunability. {\em Sci. Adv.} {\bf 7}, eabi7532 (2021).

\bibitem{Carteaux_95}V.~Carteaux, F.~Moussa and M.~Spiesser, 2D Ising-Like Ferromagnetic Behaviour for the Lamellar Cr$_2$Si$_2$Te$_6$ Compound: A Neutron Scattering Investigation. {\em Europhys. Lett.} {\bf 29}, 251 (1995).

\bibitem{Yang_23}K.~Yang, H.~Wu, Z.~Li, C.~Ran, X.~Wang, F.~Zhu, X.~Gong, Y.~Liu, G.~Wang, L.~Zhang, X.~Mi, A.~Wang, Y.~Chai, Y.~Su, W.~Wang, M.~He, X.~Yang and X.~Zhou, Spin-phonon scattering-induced low thermal conductivity in a van der Waals layered ferromagnet Cr$_2$Si$_2$Te$_6$. {\em Adv. Func. Mat}. {\bf 33}, 2302191 (2023).

\bibitem{Lin_16}M.-W.~Lin, H.~L.~Zhuang, J.~Q.~Yan, T.Z.~Ward, A.A.~Puretzky, C.M.~Rouleau, Z.~Gai, L.~Liang, V.~Meunier, B.G.~Sumpter, P.~Ganesh, P.R.C.~Kent, D.B.~Geohegan, D.Mandrus and K.~Xiao,Ultrathin nanosheets of CrSiTe$_3$: a semiconducting two-dimensional ferromagnetic material. {\em J. Mater. Chem. C}. {\bf 4}, 315 (2016).

\bibitem{Sivadas_15}N.~Sivadas, M.W.~Daniels, R.H.~Swendsen, S.~Oakamoto and 
D.~Xiao, Magnetic ground state of semiconducting transition-metal trichalcogenide monolayers. {\em Phys. Rev. B.} {\bf 91}, 235425 (2015).

\bibitem{Pei_18}Q.~Pei, X.~Wang, J.~Zou and W.~Mi, Efficient band structure modulations in two-dimensional MnPSe$_3$/CrSiTe$_3$ van der Waals heterostructures. {\em Nanotechnology}. {\bf 29}, 214001 (2018).

\bibitem{Do_22}S.-H.~Do, J.A.M.~Paddison, G.~Sala, T.J.~Williams, K.~Kaneko, K.~Kuwahara, A.F.~May, J.-Q.~Yan, M.A.~McGuire, M.B.~Stone, M.D.~Lumsden and A.D.~Christianson, Gaps in topological magnon spectra: Intrinsic versus extrinsic effects. {\em Phys. Rev. B}. {\bf 106}, L060408 (2022).

\bibitem{RESLIB}A.~Zheludev, {\em ResLib 3.4} (Oak Ridge National Laboratory, Oak Ridge, TN, 2007).

\bibitem{Nikitin_22}S.E.~Nikitin, B.~F\aa k, K.W.~Kr\"{a}mer, T.~Fennell, B.~Normand, A.M.~L\"{a}uchli and Ch.~R\"{u}egg, Thermal Evolution of Dirac Magnons in the Honeycomb Ferromagnet CrBr$_3$. {\em Phys. Rev. Lett.}. {\bf 129}, 127201 (2022).

\bibitem{Chen_21}L.~Chen, M.B.~Stone, A.I.~Kolesnikov, B.~Winn, W.~Shon, P.~Dai and J.-H.~Chung, Massless Dirac magnons in the two dimensional van der Waals honeycomb magnet CrCl$_3$. {\em 2D Mater}. {\bf 9}, 015006 (2021).

\bibitem{SPIN_W}S.~Toth and B.~Lake, Linear spin wave theory for single-Q incommensurate magnetic structures. {\em arXiv/cond-mat:1402.6069} (2014).

\bibitem{SUNNY}D.~Dahlbom, H.~Zhang, C.~Miles, S.~Quinn, A.~Niraula, B.~Thipe, M.~Wilson, S.~Matin, H.~Mankad, S.~Hahn, D.~Pajerowski, S.~Johnston, Z.~Wang, H.~Lane, Y.~W.~Li, X.~Bai, M.~Mourigal, C.D.~Batista, K.~Barros, Sunny.jl: A Julia Package for Spin Dynamics. {\em arXiv/quant-ph:2501.13095} (2025).

\bibitem{Data}T.J. Williams, D.L. Abernathy, M.D. Lumsden and A.D. Christianson,
{\em Measurements of CrSiTe$_3$ on HB3 and ARCS}, Oak
Ridge National Laboratory data repository (2026), https://doi.
org/10.14461/oncat.data/3230839.

\end{thebibliography}
\end{document}